\documentstyle[12pt]{article}

\input{psfig.sty}

\begin{document}

\begin{titlepage}


\begin{center}
\baselineskip 24pt
{\Large {\bf Possible Test for the Suggestion that Air Showers with
$E > 10^{20}$ eV are due to Strongly Interacting Neutrinos}}\\
\vspace{.5cm}
\baselineskip 14pt 
{\large Jos\'e BORDES}\\
bordes\,@\,evalvx.ific.uv.es\\
{\it Dept. Fisica Teorica, Univ. de Valencia,\\
  c. Dr. Moliner 50, E-46100 Burjassot (Valencia), Spain}\\
\vspace{.2cm}
{\large CHAN Hong-Mo, Jacqueline FARIDANI}\\
chanhm\,@\,v2.rl.ac.uk \quad faridani\,@\,hephp1.rl.ac.uk\\
{\it Rutherford Appleton Laboratory,\\
  Chilton, Didcot, Oxon, OX11 0QX, United Kingdom}\\
\vspace{.2cm}
{\large Jakov PFAUDLER}\\
jakov\,@\,thphys.ox.ac.uk\\
{\it Dept. of Physics, Theoretical Physics, University of Oxford,\\
  1 Keble Road, Oxford, OX1 3NP, United Kingdom}\\
\vspace{.2cm}
{\large TSOU Sheung Tsun}\\
tsou\,@\,maths.ox.ac.uk\\
{\it Mathematical Institute, University of Oxford,\\
  24-29 St. Giles', Oxford, OX1 3LB, United Kingdom}\\
\end{center}

\vspace{.3cm}
\begin{abstract}
The suggestion is made that air showers with energies beyond the Greisen-
Zatsepin-Kuz'min spectral cut-off may have primary vertices some 6 km
lower in height than those of proton initiated showers with energies 
below the GZK cut-off.  This estimate is based on the assumption that
post-GZK showers are due to neutrinos having acquired strong interactions
from generation-changing dual gluon exchange as recently proposed.
{\em Extremely high energy cosmic rays, cosmic
neutrinos, flavour-changing
neutral currents, duality.}
\end{abstract}

\end{titlepage}

\clearpage

Air showers at the highest known energies of around $10^{20}$ eV 
\cite{Volcano}-\cite{Auger} have long been a puzzle to cosmic ray physicists 
in that protons at such energies are thought not to be able to survive a 
long journey through the 2.7 K cosmic microwave background 
\cite{Greise,Zatsemin}, while no nearby sources are known which seem 
capable of producing such energetic particles.  Recently, following earlier
work \cite{Sigljee}-\cite{Domokos}, a suggestion was made 
that these showers may be due to neutrinos having acquired strong 
interactions at these energies \cite{Bordesetal}.  Neutrinos, being stable 
and electrically neutral, are not subject to the
Greisen-Zatsepin-Kuz'min spectral 
cut-off and can in principle reach the earth from distant sources even at 
these energies.  That they could possibly have acquired at these energies a 
strong interaction and sufficient cross section for them to initiate air 
showers is suggested by a favourite hypothesis of particle physicists that
fermion generations are a consequence of a broken gauge symmetry, which
hypothesis is in turn supported by a recent proposal that this symmetry 
may be related to dual colour \cite{Chantsou}.  If this is true, then 
the phenomenon is linked to flavour-changing neutral current hadron decays,
and estimates for their branching ratios have been derived which can serve
as tests for the hypothesis \cite{Bordesetal}.

So far, however, two things are lacking in this recent proposal: (i) an
estimate of the neutrino-air nucleus cross section showing that it is indeed
sufficient for producing air showers as observed, and (ii) a direct test
for the hypothesis with air shower data.  The purpose of this note is to 
suggest possible amendments to these deficiencies.

Strong interactions, though necessary, are in themselves not
sufficient to guarantee a large
cross section.\footnote{We are indebted to J.D.\ Bjorken for a reminder of
this fact during a talk by one of us at the Cracow Summer School in June,
1997, which started us on the following train of thought.}  If the range of 
the interaction is short, then the cross section is limited by unitarity
to a size characteristic to that range, irrespective of the strength
of the interaction.  Thus, if we were to picture the target in a
collision as a disc,
then, however strong the interaction, it cannot make the disc appear 
blacker than black.  Now, since the strong interaction of the neutrino in
the above proposal is supposedly due to the exchange of generation-changing
gauge bosons which have masses in the hundred TeV range, then the question
arises whether the neutrino will ever have enough (hadronic-sized) cross 
section with air nuclei to initiate air showers in our atmosphere.
In other words, will a nucleon in the air nucleus appear to the
neutrino as just a number of small black dots representing the partons
inside it rather than as a black disc of hadronic size?

In a general framework of generation-changing gauge bosons mediating the 
assumed new strong interactions, the answer would seem to point to the 
former alternative.  Since the mass of the new gauge bosons is bounded
below by the experimental limits on flavour-changing neutral current
decays to be in the range of 10 to several 100 TeV \cite{Cahnrari}, the
range of the interactions would seem to be only of the order of $10^{-5}$
fermi.  The nucleon will then appear to the neutrino as a collection of
very small dots and give cross sections only of the order of $10^{-12}$
barns, certainly not enough to initiate air showers.

On the other hand, if we were to accept the suggestion in \cite{Chantsou}
that generation is in fact (spontaneously broken) dual colour, 
a possibility we have already 
considered \cite{Bordesetal}, then the situation would seem to be entirely
different.  The dual gluons which are supposed to mediate the new strong
interaction between the neutrino and the partons inside the nucleon do
not represent a different degree of freedom to colour.  Indeed, in the
picture suggested in \cite{Chantsou}, the dual gluon and the gluon can
``metamorphose'' into each other.  Outside the hadron, the gluon does not 
propagate, and interactions mediated by exchanges of dual gluons will
be short-ranged.  Once inside the hadron, however, where the gluon does
propagate, the suggestion in \cite{Chantsou} was that the range of the 
interaction will be
governed by the zero gluon mass and become infinite.  The neutrino will
thus interact with the nucleon coherently and see the nucleon as a disc,
not as a collection of little black dots.  In other words, one expects 
the neutrino-nucleon cross section to be hadronic in size, and not so 
very small as in the previous scenario.

Indeed, arguing along these intuitive lines, one might even attempt a 
crude estimate of the neutrino-air nucleus cross section as follows.
Suppose that the air nucleus does appear to the neutrino as a black disc
of radius $r_A$ but that the neutrino, with yet unknown internal 
structure, appears still as a point.  Then the neutrino-nucleus cross
section is simply given as $\pi r_A^2$.  Compare this now to the 
proton-nucleus cross section.  The proton and the nucleus will appear
to each other as (almost) black discs, the proton with radius $r_p$, say.
Assuming that the proton and the nucleus will both break up as soon as
they touch, one would suggest that the proton-nucleus cross section
would be given as $\pi (r_A + r_p)^2$.  Assuming further that 
$r_A = r_p A^{1/3}$, $A$ being the atomic number of the air nucleus, which 
we take on the average to be say 15, we obtain $r_A$ to be about
2.47 $r_p$.  From this one can naively conclude that the neutrino-nucleus
cross section is about half the proton-nucleus cross section.
Although this way of estimating cross sections is admittedly crude, it
is seen to give sensible values for
proton--nucleus and proton--nucleon cross sections, with reasonable 
proton and nuclear radii, and
should thus,
we think, be good enough also for guessing the high energy neutrino-nucleus
cross section for the purpose we wish to use it.

Suppose this is true.  We conclude first that neutrinos at these energies
will have enough cross section to initiate air showers, and secondly,
since the cross section is smaller than for protons, the neutrino will
be somewhat more penetrating and initiate air showers at lower altitudes
on the average.  The second fact, we believe, may be used as a criterion
to distinguish neutrino showers statistically from proton showers and hence 
test the original suggestion that the highest energy showers are initiated 
by neutrinos rather than protons.

It is not difficult to make our statement above more quantitative.  Air
density varies with height $h$ in cm above sea-level roughly as:
\begin{equation}
\rho(h) = 1.2 \ (\exp -h/h_0) \times 10^{-3} {\rm gm/cm}^3,
\label{rhoh}
\end{equation}
with the attenuation length:
\begin{equation}
h_0 = 7.6 \times 10^5 {\rm cm}.
\label{h0}
\end{equation}
Suppose the flux of a particle has initial value $f_{inc}$.  Let
$\theta$ be the angle to the zenith at the point the shower axis hits
the earth's surface and $x$ the distance from this point measured
aloong the shower axis.  Then the flux, after penetrating to the point
$(x,\theta)$, will be attenuated to the value:
\begin{equation}
f(x,\theta) = f_{inc} \exp \left\{ K(\sigma) \int_{\infty}^x dx' 
   \rho(h(x',\theta)) \right\},
\label{fxtheta}
\end{equation}
where the height $h$ expressed in terms of $x$ and $\theta$ is:
\begin{equation}
h = \sqrt{R^2 + 2 x R cos\theta + x^2} - R,
\label{hxtheta}
\end{equation}
with $R$ being the radius of the earth.  The attenuation constant $K$ is:
\begin{equation}
K(\sigma) = (N/A) \sigma,
\label{Ksigma}
\end{equation}
where $N$ is the Avogadro number, $A$ the atomic number of the air nucleus,
and $\sigma$ the incident particle-nucleus cross section.  For protons,
$K^{-1}$ is about 60 gm/cm$^2$ at these high energies, and if we were right in
our estimate above, $K$ would be about one half of this value for neutrinos.

\begin{figure}[htb]
\centerline{\psfig{figure=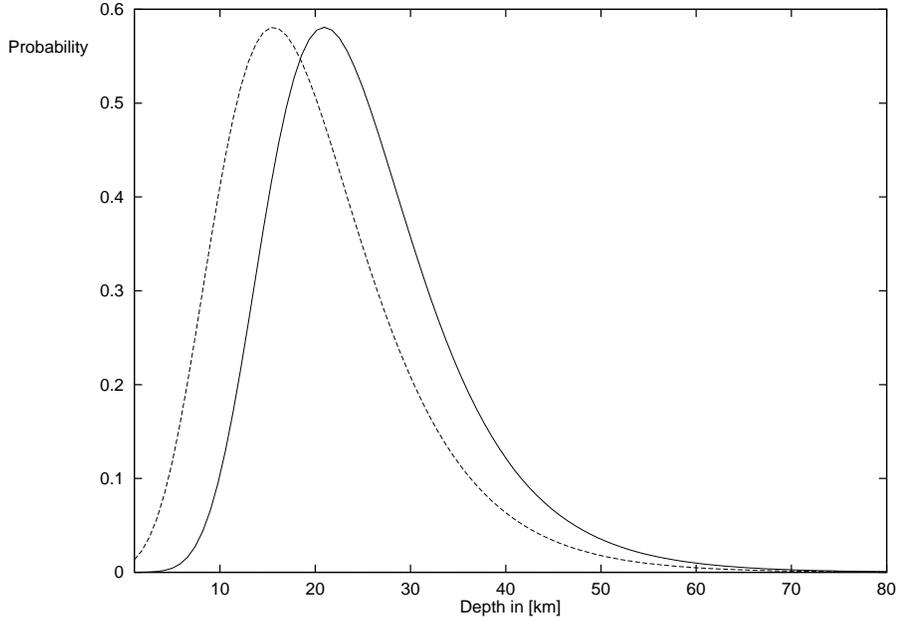,width=0.6\textwidth}}
\vspace{-1cm} 
\caption{Probability distributon (arbitrary units) of primary vertices for proton-initiated
(full curve) and neutrino-initiated (dotted curve) air showers.}
\label{penprob}
\end{figure}

The probability for effecting a collision and producing an air shower at
$x$ and $\theta$ is then:
\begin{equation}
F(x, \theta) = K(\sigma) \rho(h(x,\theta)) f(x, \theta).
\label{Fxtheta}
\end{equation}
This, being a product of two exponentials, one decreasing and the other 
increasing with height, has a maximun at some $x$ which will then 
be the most likely place where an air shower will be initiated.  In 
Figure \ref{penprob}, we show the distribution function $F$ of the 
``primary vertex'' for respectively proton- and neutrino-initiated showers
as a function of $x$ at $\theta = 0$, i.e.\ vertically down.  One sees that
the maxima for protons and neutrinos differ by around 6 km in height,
with proton showers occuring at around 21 km and neutrino showers at
around 15 km.  
\vspace{0.7cm}   

We conclude therefore that if, as suggested, showers below the GZK cut-off
are mostly proton-initiated while those above the GZK cut-off are 
neutrino-initiated, then the primary vertices of those below GZK should 
cluster around 21 km in height while those above GZK should cluster at 
around 15 km.\footnote{This assumes that detection efficiency has been
folded in.}  The maxima in both distributions being quite sharp, as seen
in Figure \ref{penprob}, the clusters should be well-separated from one
another.  

\begin{figure}[htb]
\centerline{\psfig{figure=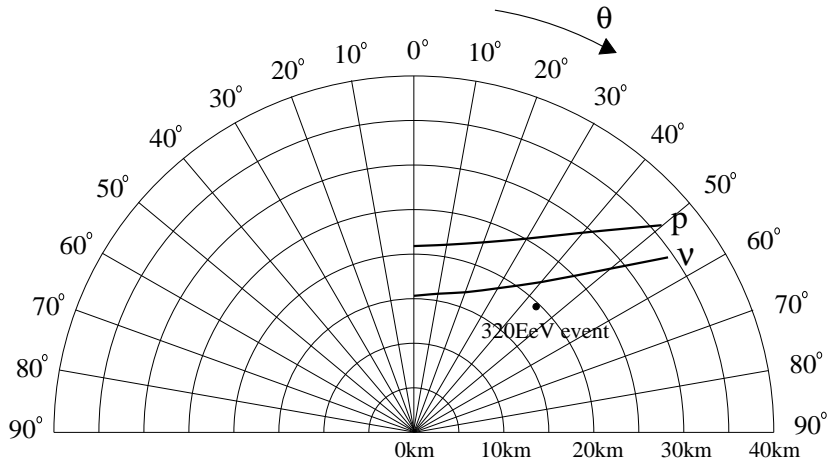,width=0.8
\textwidth}}
\vspace{1cm}
\caption{The positions of the distribution maxima for varying $\theta$}
\label{maxangle}
\end{figure}

The calculation can be repeated for all incident angles $\theta$ giving
very similar distributions, although the maximum and also the width
of the maximum will depend on $\theta$.  In Figure \ref{maxangle}, we
plot the positions of the distribution maxima for varying $\theta$,
for both the proton and the neutrino.  One sees that
the two curves are well-separated  
with the neutrino curve lying much lower than the proton curve.  If we
take each event and plot the position of its primary vertex on 
Figure \ref{maxangle}, the prediction is that pre-GZK events representing
proton showers will cluster around the top curve while post-GZK events
representing neutrino showers will cluster around the bottom curve, 
with a clear separation between them.
\vspace{1cm}

We recognize that the primary vertex is in most experiments difficult or
perhaps even impossible to determine accurately.  But in a detector like 
the Fly's Eye \cite{Flyseye}, the development profile of the shower is measured,
and by examining the profile function closely near the beginning one may get a
reasonable idea of where the primary vertex is located.  As an exercise, 
we take the development profile of the highest energy shower known at 3.2
$\times 10^{20}$ eV detected by Fly's Eye and look for the point where 
fluorescence was first detected, which was at a depth of around 200 gm/cm$^2$. 
This corresponds to a vertical height of around 12 km or to $x=19.5$
km for the observed $\theta=43^0.9$. 
If we boldly call this the primary vertex and plot it on 
Figure \ref{maxangle}, 
we obtain the point shown.  From Figure
\ref{penprob} 
we see that the 
probability of a proton shower having its primary vertex at or lower than 
12 km is only about 5 percent, which means that, other things being equal 
and taking this information at its face value, it would seem that this 
event is much more likely to be from a neutrino as suggested in 
\cite{Bordesetal} than from a proton.  We realize, of course, that we
have been extremely naive to identify the primary vertex as the point 
when light first shows in the Fly's Eye detector, which identification
should have been made only by the experimenters themselves after a 
careful analysis of the shower development profile, the detection 
efficiency etc.  For all we know, the shower might have 
started much higher up without showing any light.  However, as far as 
the method is concerned, it would seem that, given the development
profiles of two showers with primary vertices differing by as much as
6 km in height, there should be no difficulty in distinguishing them.
It appears to us therefore that with the data collected by Fly's Eye, it 
may already be possible to decide whether the suggestion is feasible.  
In any case, for the Auger project \cite{Auger} which has also the Fly's 
Eye's facility, only better, it seems that with some effort, it ought 
to be a relatively simple matter.

If such a separation is indeed seen in experiment, then it would be a 
rather good test of the hypothesis that pre- and post-GZK showers are
initiated by different particles with different cross sections.  In view
of the absence of any other stable particles known, with hadronic yet
somewhat smaller cross section than the proton, it would seem then that 
there is a fair chance of the latter being initiated by neutrinos.  The
converse, however, would be harder to conclude if no clear difference in 
height is seen since the neutrino cross section used in the analysis above
has been so crudely estimated.  Nevertheless, it seems to us an attempt 
worth making since the prize is so attractive.

The crude picture outlined in the beginning for high energy neutrino
interactions suggests in fact also some differences in the development
of showers due respectively to neutrinos and to protons.  The neutrino in 
this picture being elementary and the proton composite, it seems that the 
development profile of neutrino-initiated showers would differ from that
that of proton-initiated showers in much the same way that showers initiated 
by nuclei differ from those initiated by protons.  However, the average 
number of partons in the proton being probably small compared with the
number of nucleons in a (say iron) nucleus, the difference would be less
marked and we are not sure it would be noticeable.  We think that the
difference in height of the ``primary vertex'' as described above would
be a more hopeful means for differentiating the two primaries.

Looking further, suppose we are convinced by further analysis based on the 
above method or otherwise that air showers beyond the GZK cut-off are 
indeed due to neutrinos.  Then by turning the argument around, we might 
imagine using the Auger project \cite{Auger} as an apparatus for measuring the 
high energy neutrino cross section.  For example, if we draw the contours of
the type shown in Figure \ref{maxangle}, one for each value of $\sigma$, 
then by plotting each event observed above the GZK cut-off in the figure
and seeing on which contour it lies, we obtain for it some value of
$\sigma$.  If we next plot the number of post-GZK events against $\sigma$,
we shall be able to read off directly the neutrino-nucleus cross section
from the position of the peak of the distribution.

Going further still, we might even imagine using the Auger project as a
spectrometer for studying the mass spectrum of generation-changing gauge 
and Higgs bosons.  In the incoming neutrino beam, there will be presumably
also anti-neutrinos, and if generation-changing bosons do exist, then
an anti-neutrino on hitting an electron present in the atmosphere can 
form one of these bosons provided that the collision occurs at the right
energy.  The highest shower known at present has $E = 3.2 \times
10^{20}$ eV
corresponding in a collision with an electron to a C.M. energy of around 
18 TeV, which is not far from the estimates for the masses of the lowest
generation-changing Higgs bosons obtained from the dual scheme 
\cite{Chantsou,Bordesetaln}.  Should the spectrum for cosmic ray neutrinos
extend further up, and at the moment we do not know any reason why it
should not, then the Auger project should be able to sweep the mass 
region from 10 TeV upwards and see generation-changing bosons occuring as 
resonance peaks in a manner similar to that in ordinary spectroscopy 
experiments.

\vspace{.5cm}
\noindent {\large {\bf Acknowledgement}}\\
We are much indebted first to Jeremy Lloyd-Evans for introducing us to
this fascinating subject of high energy air showers, and then to both
him and Alan Watson for supplying us with valuable information and advice
during the course of this short piece of work.  We thank also Andrzej
Bialas for correcting an error.
JF would like to thank
the Rutherford Appleton Laboratory for hospitality, and JP gratefully
aknowledges financial support from the German Academic Exchange
Service (DAAD).

\end{document}